\documentclass[prd, 11pt,nofootinbib, superscriptaddress, preprintnumbers,floatfix]{revtex4}
\usepackage{amsmath,amssymb}
\usepackage{graphicx}
\usepackage{subfigure}
\usepackage{hyperref}
\usepackage{nicefrac}
\usepackage{color}

\newcommand{\mpi}{M_{\pi^\pm}}
\newcommand{\mpin}{M_{\pi^0}}
\newcommand{\mpios}{M_\pi^\mathrm{OS}}

\newcommand{\tr}{\mathrm{Tr}}

\newcommand{\im}{\mathrm{Im}}

\graphicspath{{plots/}}

\begin{document}

\title{Isospin-0 $\pi\pi$ s-wave scattering length from twisted mass lattice QCD}

\newcommand\bn{Helmholz-Institut f\"ur Strahlen- und Kernphysik and Bethe Center for Theoretical
             Physics, Universit\"at Bonn, D-53115 Bonn, Germany}
\newcommand\rom{Dipartimento di Fisica, Universit{\`a} and INFN di Roma Tor Vergata, 00133 Roma, Italy}
\newcommand\bern{Albert Einstein Center for Fundamental Physics, University of Bern, 3012 Bern, Switzerland}
\newcommand\cypa{Computation-based Science and Technology Research Center, The Cyprus Institute, PO Box 27456, 1645 Nicosia, Cyprus}
\newcommand\cypb{Department of Physics, University of Cyprus, PO Box 20537, 1678 Nicosia, Cyprus}
\newcommand\wup{Fakult\"at f\"ur Mathematik und Naturwissenschaften, Bergische Universit\"at Wuppertal, 42119 Wuppertal, Germany}
\newcommand\roma{Centro Fermi - Museo Storico della Fisica e Centro Studi e Ricerche
Enrico Fermi, Compendio del Viminale, Piazza del
Viminiale 1, I-00184, Rome, Italy}
\newcommand\romb{Dipartimento di Fisica, Universit{\`a} di Roma ``Tor Vergata",
Via della Ricerca Scientifica 1, I-00133 Rome, Italy}

\author{L.~Liu}\email{liuming@hiskp.uni-bonn.de}\affiliation{\bn}
\author{ S.~Bacchio}\affiliation{\cypb}\affiliation{\wup}
\author{P.~Dimopoulos}\affiliation{\roma}\affiliation{\romb}
\author{J.~Finkenrath}\affiliation{\cypa}
\author{R.~Frezzotti}\affiliation{\rom}
\author{C.~Helmes}\affiliation{\bn}
\author{C.~Jost}\affiliation{\bn}
\author{B.~Knippschild}\affiliation{\bn}
\author{B.~Kostrzewa}\affiliation{\bn}
\author{H.~Liu}\affiliation{\bern}
\author{K.~Ottnad}\affiliation{\bn}
\author{M.~Petschlies}\affiliation{\bn}
\author{C.~Urbach}\email{urbach@hiskp.uni-bonn.de}\affiliation{\bn}
\author{M.~Werner}\affiliation{\bn}
\collaboration{ETM Collaboration}

\begin{abstract}
  \begin{center}
    \includegraphics[draft=false,width=.18\linewidth]{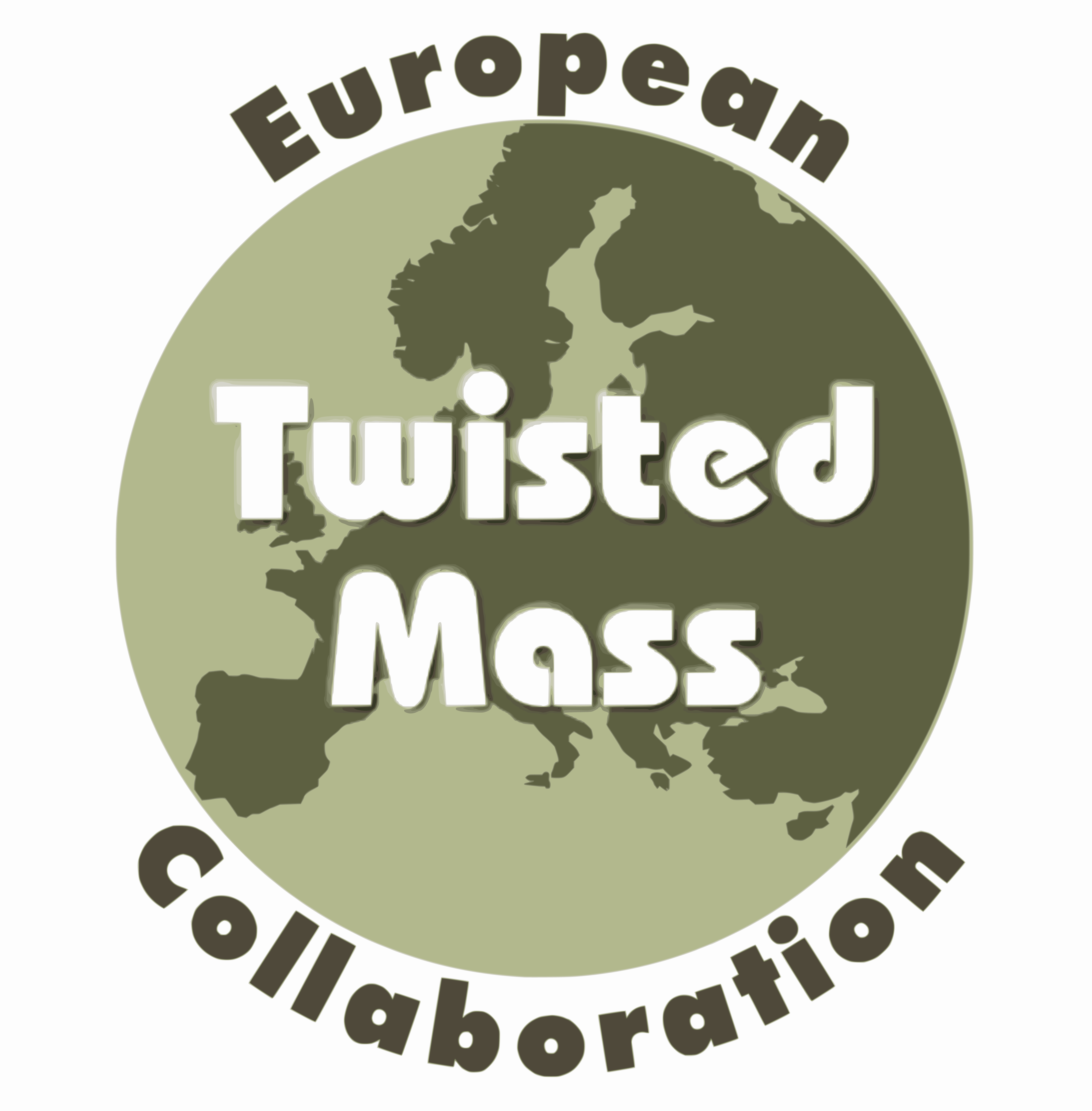}
  \end{center}
  \vspace*{0.25cm}

We present results for the isospin-0 $\pi\pi$ s-wave scattering length
calculated with Osterwalder-Seiler valence quarks on Wilson twisted
mass gauge configurations. We use three $N_f = 2$ ensembles with
unitary (valence) pion mass at its physical value (250~MeV), at
240~MeV (320~MeV) and at 330~MeV (400~MeV),
respectively. By using the stochastic Laplacian Heaviside quark
smearing method, all quark propagation diagrams contributing to the
isospin-0 $\pi\pi$ correlation function are computed with sufficient
precision. The chiral extrapolation is performed to obtain the  
scattering length at the physical pion mass. Our result $M_\pi
a^\mathrm{I=0}_0 = 0.198(9)(6)$ agrees reasonably well with various
experimental measurements and theoretical predictions. Since we only
use one lattice spacing, certain systematics uncertainties, especially
those arising from unitary breaking, are not controlled in our result. 

\end{abstract}

\maketitle

\clearpage

 \section{Introduction}

Quantum chromodynamics (QCD) is established as
the fundamental theory of the strong interactions. 
QCD at low energies is largely determined by chiral symmetry, which is
spontaneously broken. The effective theory of QCD at low energies is
chiral perturbation theory
($\chi$PT)~\cite{Weinberg:1978kz,Gasser:1983yg,Gasser:1987ah},
representing a systematic expansion in the quark masses and
momenta. Elastic $\pi\pi$ scattering provides an ideal testing ground for the mechanism of
chiral symmetry breaking. Since only the pions
-- the pseudo-Goldstone bosons of SU$(2)$ chiral symmetry -- are involved, 
the expansion is expected to converge rapidly.
In fact, the s-wave scattering length in the weakly repulsive isospin-2 channel
can be reproduced by leading order (LO) $\chi$PT ~\cite{Weinberg:1966kf}
with a deviation of only 0.5\% when compared to the results obtained from
experiments combined with Roy
equations~\cite{Batley:2010zza}.

However, in the isospin-0 channel the situation is different: the interaction is
attractive and much stronger than in the isospin-2 channel. The agreement
between LO $\chi$PT and experiments for the s-wave scattering length in the
isospin-0 channel is much less impressive: they deviate by around 30\%
~\cite{Weinberg:1966kf, Batley:2010zza, Pislak:2003sv}. Moreover,
this channel accommodates the lowest QCD resonance -- the
mysterious $\sigma$ or $f_0(500)$ scalar meson. Although it plays a
crucial role in some fundamental features of QCD, its existence was
disputed for a long time. Only recently it was established unambiguously with
dispersive analyses and new experimental data, see
Ref.~\cite{Pelaez:2015qba} for a review.

This makes a nonperturbative, {\it ab initio} computation of $\pi\pi$
interaction properties in the isospin-0 channel directly from QCD highly
desirable. Lattice QCD is the only available method to perform such
a computation with controlled systematic uncertainties.  L\"uscher
showed that the infinite volume scattering parameters can be
related to the discrete spectrum of the eigenstates in a finite-volume 
box~\cite{Luscher:1986pf, Luscher:1990ux}. This allows one to compute scattering
properties in lattice QCD, which is necessarily implemented in a
finite volume and Euclidean space-time. 

For the isospin-2 channel, many lattice results have become
available. See
Refs.~\cite{Yagi:2011jn,Fu:2013ffa,Sasaki:2013vxa,Helmes:2015gla} for the most recent ones.
For the isospin-0 channel the situation is more complicated mainly due
to the  fermionic disconnected diagrams contributing here, which are
challenging to compute in lattice QCD. 
To date there are only two full lattice QCD computations dedicated to this
channel ~\cite{Fu:2013ffa, Briceno:2016mjc}. In
Ref.~\cite{Fu:2013ffa}, the s-wave scattering length was computed for
three unphysically large pion masses. An extrapolation to the
physical point was performed to obtain the scattering length at
physical pion mass. The authors of Ref.~\cite{Briceno:2016mjc} on the
other hand extracted many energy
eigenstates in the corresponding channel and obtained the scattering
amplitudes for two values of pion mass -- 236~MeV and 391~MeV. The
information about the $\sigma$ meson is deduced from the pole
structure in the scattering amplitudes at the two unphysical pion
masses, respectively. 
 
In this work we compute the scattering length of the isospin-0
$\pi\pi$ channel in twisted mass lattice QCD~\cite{Frezzotti:2000nk}
and L\"uscher's finite volume method~\cite{Luscher:1986pf,
  Luscher:1990ux}. As discussed in Ref.~\cite{Buchoff:2008hh}, the
explicit isospin breaking of the twisted mass quark action makes it
prohibitively complicated to study this channel with this
action. To circumvent this complication we use a mixed
  action approach with Osterwalder-Seiler
  quarks~\cite{Frezzotti:2004wz} in the valence 
sector, which preserves isospin symmetry. This approach introduces
additional lattice artefacts due to unitarity breaking. These  
lattice artefacts are of $\mathcal{O}(a^2)$ and will vanish only in
the continuum limit. In particular, due to isospin breaking in the sea
there is possibly residual mixing with $I=2, I_z=0$.
Since we use only one value of lattice spacing,
systematic uncertainties in our results are not fully
controlled. Further calculations are needed to explicitly address
these uncertainties. However, they are beyond the scope of this work.

This paper is organized as follows. The lattice setup is discussed
in Sec.~\ref{sec:actions}. L\"uscher's finite volume method is
introduced in Sec.~\ref{sec:FiniteVolumeMethod}. In
Sec.~\ref{sec:FiniteVolumeSpectrum} we present the computation of
the finite volume spectrum of the isospin-0 $\pi\pi$ system. The
result for the scattering length is given in
Sec.~\ref{sec:results}. The last section is devoted to a brief
summary and discussions.  

\section{Lattice action}
\label{sec:actions}

\begin{table}[t!]
 \centering
 \begin{tabular*}{.9\textwidth}{@{\extracolsep{\fill}}lccccc}
  \hline\hline
  ensemble & $\beta$ & $c_{\mathrm{sw}}$ &$a\mu_\ell$  &$(L/a)^3\times T/a$ & $N_\mathrm{conf}$  \\ 
  \hline\hline
  $cA2.09.48$ &2.10 &1.57551 &0.009  &$48^3\times96$ & $615$ \\
  $cA2.30.48$ &2.10 &1.57551 &0.030  &$48^3\times96$ & $352$ \\
  $cA2.60.32$ &2.10 &1.57551 &0.060  &$32^3\times64$ & $337$ \\
  \hline\hline
 \end{tabular*}
 \caption{The gauge ensembles used in this study. The labeling of
   the ensembles follows the notations in
   Ref.~\cite{Abdel-Rehim:2015pwa, Baron:2010bv}. In addition to the relevant input
   parameters we give the lattice volume $(L/a)^3\times T/a$ and  the number of evaluated
   configurations $N_\mathrm{conf}$.}
 \label{tab:setup}
\end{table}

The results presented in this paper are based on the gauge
configurations generated by the European Twisted Mass Collaboration
(ETMC) with Wilson clover twisted mass quark action at maximal  
twist~\cite{Frezzotti:2000nk}. The gauge action is the Iwasaki gauge
action~\cite{Iwasaki:1985we}. 
We use three $N_f = 2$ ensembles with
pion mass at the physical value, at 240~MeV and at 330~MeV,
respectively. The lattice spacing is $a=0.0931(2)\ \mathrm{fm}$ forflavor
all three ensembles, as found in Ref.~\cite{Abdel-Rehim:2015pwa} up to $\mathcal{O}(a^2)$ lattice artefacts.  
In Table~\ref{tab:setup} we list the three
ensembles with the relevant input parameters, the lattice volume and
the number of configurations. More details about the ensembles are
presented in Ref.~\cite{Abdel-Rehim:2015pwa}. 

The sea quarks are described by the Wilson clover twisted mass
action. The Dirac operator for the light quark doublet consists of the
Wilson twisted mass Dirac operator~\cite{Frezzotti:2000nk} combined
with the clover term (in the so-called physical basis)
\begin{equation}
  \label{eq:Dlight}
  D_\ell = \widetilde{\nabla} -i\gamma_5\tau_3 \left[W_\mathrm{cr} + 
  \frac{i}{4}c_\mathrm{sw}\sigma^{\mu\nu}\mathcal{F}^{\mu\nu}\right] + \mu_\ell\,,
\end{equation}
with $\widetilde{\nabla} = \gamma_\mu(\nabla^\ast_\mu+\nabla_\mu)/2$, $\nabla_\mu$ and
$\nabla_\mu^\ast$ the forward and backward lattice covariant
derivatives. Here $c_{sw}$ is the so-called
Sheikoleslami-Wohlert improvement coefficient
\cite{Sheikholeslami:1985ij} multiplying the clover term
and $W_\mathrm{cr} = -ra\nabla_\mu^\ast\nabla_\mu +
m_\mathrm{cr}$, with $m_\mathrm{cr}$ the critical mass. $\mu_\ell$ is
the average up/down (twisted) quark mass. $a$ is the lattice spacing
and $r=1$ the Wilson parameter. $D_\ell$ acts on a flavor doublet
spinor $\psi_\ell = (u,d)^T$.  In our case the clover term is not used for $\mathcal{O}(a)$
 improvement but serves to
significantly reduce the effects of isospin
breaking~\cite{Abdel-Rehim:2015pwa}. 

The critical mass has been determined as described in
Refs.~\cite{Chiarappa:2006ae,Baron:2010bv}. This guarantees
automatic $\mathcal{O}\left(a\right)$ improvement
\cite{Frezzotti:2003ni}, which is one of the main advantages of the
Wilson twisted mass formulation of lattice QCD. 

In the valence sector we introduce quarks in the so-called
Osterwalder-Seiler (OS) discretization~\cite{Frezzotti:2004wz}. The 
corresponding \textit{up} and \textit{down} single flavor lattice
Dirac operators read 
\begin{equation}
  \label{eq:DOS}
  D_\ell^\pm = \widetilde{\nabla} \pm i\gamma_5 \left[W_\mathrm{cr} + 
  \frac{i}{4}c_\mathrm{sw}\sigma^{\mu\nu}\mathcal{F}^{\mu\nu}\right] +
  \mu_\ell^\mathrm{OS}\,.
\end{equation}
From this definition it is apparent that OS \textit{up} and
\textit{down} quarks have explicit SU$(2)$ isospin symmetry if for
both e.g. $D_\ell^+$ was used. The
matching of OS to unitary actions is performed by matching the quark
mass values, i.e. $\mu_\ell^\mathrm{OS} = \mu_\ell$. The value of
$m_\mathrm{cr}$ in the OS action can be shown to be identical to the
unitary one and $\mathcal{O}(a)$ improvement stays
valid~\cite{Frezzotti:2004wz}. Moreover, we have shown in
Ref.~\cite{Ottnad:2015hva} that in such a mixed action approach
disconnected contributions to $\eta$ and $\eta'$ mesons can be
computed and the results agree with the unitary
ones~\cite{Michael:2013gka} in the continuum limit. Therefore, 
 this mixed action approach should works also in the case of
$\pi\pi$ scattering, where disconnected contributions can be expected
to be less important, since OZI suppression is in place. 
However, there is a potential complication arising from
  the double poles in flavor-neutral meson propagators present in a
  quenched or partially quenched theory~\cite{Bernard:1994zp}. 
The scalar correlators with disconnected diagrams suffer from
unphysical contributions due to the double poles. 
The unphysical contributions to the $a_0$ and $\pi\pi$ correlators have been studied in 
Refs.~\cite{Aubin:2008wk, Bardeen:2001jm, Golterman:2005xa, Bernard:1995ez}. 
In this work, we are not going to consider this problem since the formula of these unphysical
contributions for our partially quenched approach is not available. Also, as will be presented in Sec. 
~\ref{sec:FiniteVolumeSpectrum}, our numerical results do not indicate
large unphysical contributions. All the correlators we computed
numerically are positive within the obtained statistics and are well
described by a single exponential function of $t/a$ in a reasonably large
time range. This would not be the case if there were large unphysical
contributions as shown in Refs.~\cite{Aubin:2008wk, Bardeen:2001jm,
  Golterman:2005xa, Bernard:1995ez}. Nevertheless, one should keep in
mind that the effects of the double poles  
may cause uncertainties that are not considered in our results.

Masses computed with OS valence quarks differ from those computed with
twisted mass valence quarks by lattice artefact of
$\mathcal{O}(a^2)$, in particular
\[
(M_\pi^\mathrm{OS})^2 - (M_\pi)^2\ =\ \mathcal{O}(a^2)\,.
\]
For twisted clover fermions this difference is much reduced as
compared to twisted mass fermions~\cite{Abdel-Rehim:2015pwa}, however,
the effect is still sizable. We use the OS pion mass in this paper,
with the consequence that the pion mass value of the cA2.09.48
ensemble takes a value larger (around 250~MeV) than the physical one. 

As a smearing scheme we use the stochastic Laplacian Heavyside (sLapH)
method~\cite{Peardon:2009gh,Morningstar:2011ka} for our
computation. The details of the sLapH parameter choices for a set of
$N_f = 2+1+1$ Wilson twisted mass ensembles are given in
Ref.~\cite{Helmes:2015gla}. The parameters for the ensembles used in
this work are the same as those for $N_f = 2+1+1$ ensembles with the
corresponding lattice volume.

\section{L\"uscher's finite volume method}
\label{sec:FiniteVolumeMethod}

L\"uscher's finite volume method provides a direct relation between
the energy eigenvalues of a two-particle system in a finite box and
the scattering phase shift of the two particles in the infinite
volume. Considering two particles with mass $m_1$ and $m_2$ in a cubic
box of size $L$, the energy of this system in the center-of-mass frame
reads
\begin{equation}
  \label{eq:ScatMomentum}
  E = \sqrt{m_1^2 + \vec{k}^2} + \sqrt{m_2^2 + \vec{k}^2}\,,
\end{equation}
where $\vec{k}$ is the scattering momentum. For the following
discussion, it is convenient to define a dimensionless variable $q$
via  
\begin{equation}
  q^2 = \frac{\vec{k}^2 L^2 }{ (2\pi)^2}\,,
\end{equation}
which differs from an integer due to the interaction of the two
particles.

The general form of L\"uscher's formula reads~\cite{Luscher:1990ux}:
\begin{equation}
  \label{eq:luscher}
  \det \left [ e^{2i \delta_l} \delta_{ll^\prime} \delta_{nn^\prime} -
    \frac{\mathcal{M}^{\Gamma}_{ln,l^\prime n^\prime} +
      i}{\mathcal{M}^{\Gamma }_{ln,l^\prime n^\prime}- i} \right ] =
  0,  
\end{equation}
where $\delta_l$ is the phase shift of the partial wave with angular
momentum $\{l\}$, $\Gamma$ denotes an irreducible representation (irrep) of
the cubic group. The matrix in the determinant is labeled by the pair
$(l, n)$, in which $\{l\}$ are the angular momenta subduced into the irrep
$\Gamma$ and $n$ is an index indicating the $n^{th}$ occurrence of that
$l$ in the irrep. The matrix element
$\mathcal{M}^{\Gamma}_{ln,l^\prime n^\prime}$ is a complicated 
function of $q$ but can be computed numerically.  

In this work we are interested in the s-wave low energy scattering in
the isospin-0 $\pi \pi$ channel. Therefore, we will compute only the lowest
energy level in the center-of-mass frame. In this case one
should consider the irrep $A_1^{+}$. Assuming that the effects of the partial
waves with $l \geq 4$ are negligible, the matrix in the determinant of
Eq.~\ref{eq:luscher} becomes one-dimensional and the equation reduces to
\begin{equation}
  \label{eq:luscherA1}
  q \cot \delta_0 (k) = \frac{\mathcal{Z}_{00}(1; q^2)}{\pi^{3/2}}\,,
\end{equation}
where $\mathcal{Z}_{00}(1; q^2)$ is the L\"uscher zeta-function which
can be evaluated numerically for given value of $q^2$. Using the
effective range expansion of s-wave elastic scattering near 
threshold, we have 
\begin{equation}
  \label{eq:effrange_expansion}
  k \cot \delta_0 (k) = \frac{1}{a_0} + \frac{1}{2} r_0 k^2 + \mathcal{O}(k^4)\,,
\end{equation}
where $a_0$ is the scattering length and $r_0$ is the effective
range parameter. Once the finite volume energy $E$ is determined from
lattice QCD simulations, the scattering length $a_0$ can be calculated
from the following relation
\begin{equation}
  \label{eq:ScatLen}
  \frac{2\pi}{L} \frac{\mathcal{Z}_{00}(1; q^2)}{\pi^{3/2}} =
  \frac{1}{a_0} + \frac{1}{2} r_0 k^2 + \mathcal{O}(k^4)\,.
\end{equation}
It should be pointed out that L\"uscher's formulas presented here,
i.e. Eqs.~\ref{eq:luscher} and \ref{eq:luscherA1}, are for the
scattering processes with $k^2 > 0$. The phase shift $\delta_0 (k)$ in
the continuum is only defined for positive $k^2$. In the case of
negative $k^2$, one can introduce a phase $\sigma_0(k)$ which is
related to $\delta_0(k)$ by analytic continuation of $\tan \sigma_0(k) =
-i \tan \delta_0(k)$~\cite{Luscher:1990ux}. Only when there is a bound
state at the corresponding energy, the phase $\sigma_0(k)$ has
physical interpretation and its value equals to $-\pi/4$ (modulo
$\pi$) in the continuum and infinite volume limit. For the purpose of this
paper -- calculating the scattering length, we will only consider the
lowest energy level in the center-of-mass frame. Since the interaction
in the isospin-0 $\pi\pi$ channel is attractive, this energy level is
below the threshold, i.e. $k^2 < 0$. For convenience, in the following
we will always use the notation $k\cot \delta_0(k)$, which is understood 
as the analytically continued form for $k^2<0$. 
Please note that Eq.~\ref{eq:ScatLen} holds for both positive
and negative $k^2$ as long as the modulus of $k^2$ is close to zero.

\section{Finite volume spectrum}
\label{sec:FiniteVolumeSpectrum}

In lattice QCD, the discrete spectra of hadronic states are extracted
from the correlation functions of the interpolating operators that
resemble the states. Due to the isospin symmetry breaking of the
twisted mass quark action, it is difficult to investigate the
isospin-0 $\pi\pi$ channel directly in unitary twisted mass lattice
QCD~\cite{Buchoff:2008hh}. For this reason we use a mixed action
approach with the OS action~\cite{Frezzotti:2004wz} in the valence
sector and choose $D_\ell^+$ in Eq.~\ref{eq:DOS} for both {\it up} and
{\it down} quarks, so that the isospin symmetry is guaranteed in
the valence sector. In this section we describe our methodology to
calculate the energies of the isospin-0 $\pi\pi$ system.  

\subsection{Computation of the correlation functions}

We define the interpolating operator that represents the isospin-0
$\pi\pi$ state in terms of OS valence quarks
\begin{equation}
  \label{Eq:pipioperator}
  \mathcal{O}_{\pi\pi}^\mathrm{I=0}(t) = \frac{1}{\sqrt{3}}(
  \pi^{+}\pi^{-}(t)\ +\ \pi^{-}\pi^{+}(t)\ +\
  \pi^{0}\pi^{0}(t))\,, 
\end{equation}
with single pion operators summed over spatial coordinates
$\mathbf{x}$ to project to zero momentum 
\begin{equation}
  \label{eq:piop}
  \begin{split}
    \pi^{+}(t) = \sum_\mathbf{x} \bar{d} \gamma_5 u
    (\mathbf{x},t)\,, \quad &
    \pi^{-}(t) = \sum_\mathbf{x} \bar{u} \gamma_5 d (\mathbf{x},t)\,, \\
    \pi^{0}(t) = \sum_\mathbf{x} \frac{1}{\sqrt{2}} & (
    \bar{u}\gamma_5 u - \bar{d}\gamma_5 d)(\mathbf{x},t)\,.
  \end{split}
\end{equation}
Here $u$ and $d$ represent the OS {\it up} and {\it down} quarks,
respectively. With OS valence quarks all three pions are mass
degenerate and will be denoted as $\mpios$. 

\begin{figure}[t]
  \includegraphics[width=0.25\linewidth]{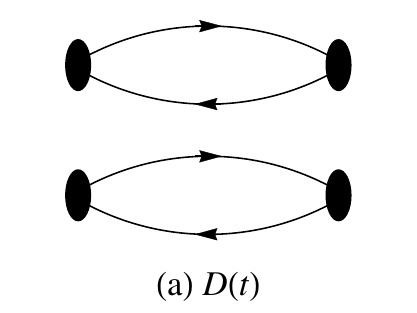}\includegraphics[width=0.25\linewidth]{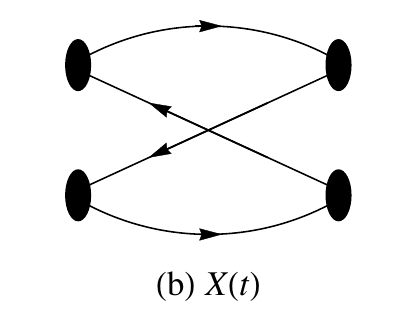}\includegraphics[width=0.25\linewidth]{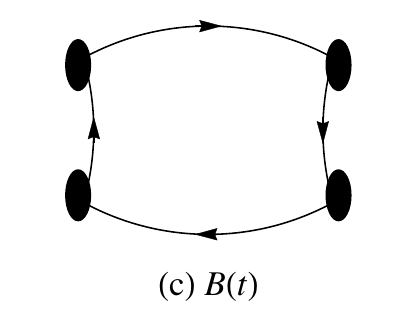}\includegraphics[width=0.25\linewidth]{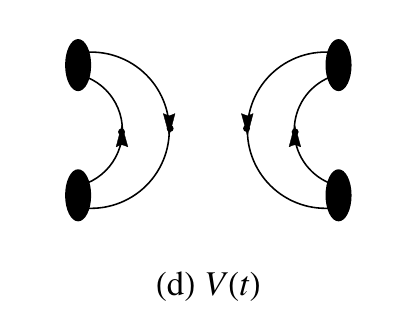} 
  
  \includegraphics[width=0.25\linewidth]{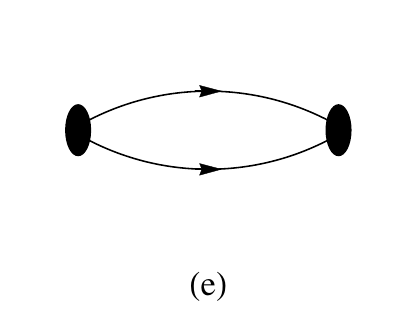}\includegraphics[width=0.25\linewidth]{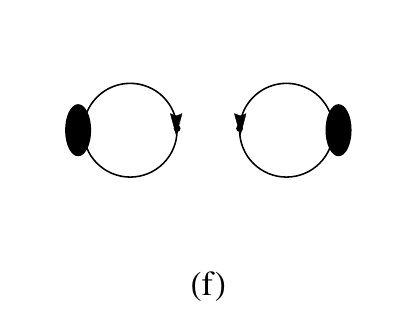}\includegraphics[width=0.25\linewidth]{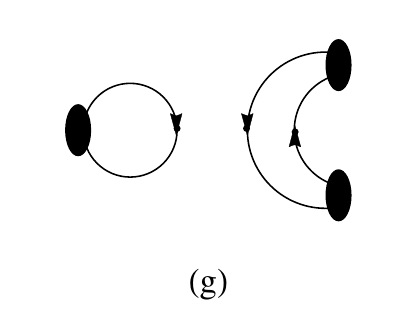} 
    \caption{Diagrams contributing to the correlation
    functions. (a)--(d) represent the usual contributions to
    $C_{\pi\pi}$, while (e)--(f) need to be taken into account due to
    unitarity breaking effects.}
  \label{fig:diagrams}
\end{figure}

\begin{figure}[t]
  \centering
  \includegraphics[width=0.5\linewidth]{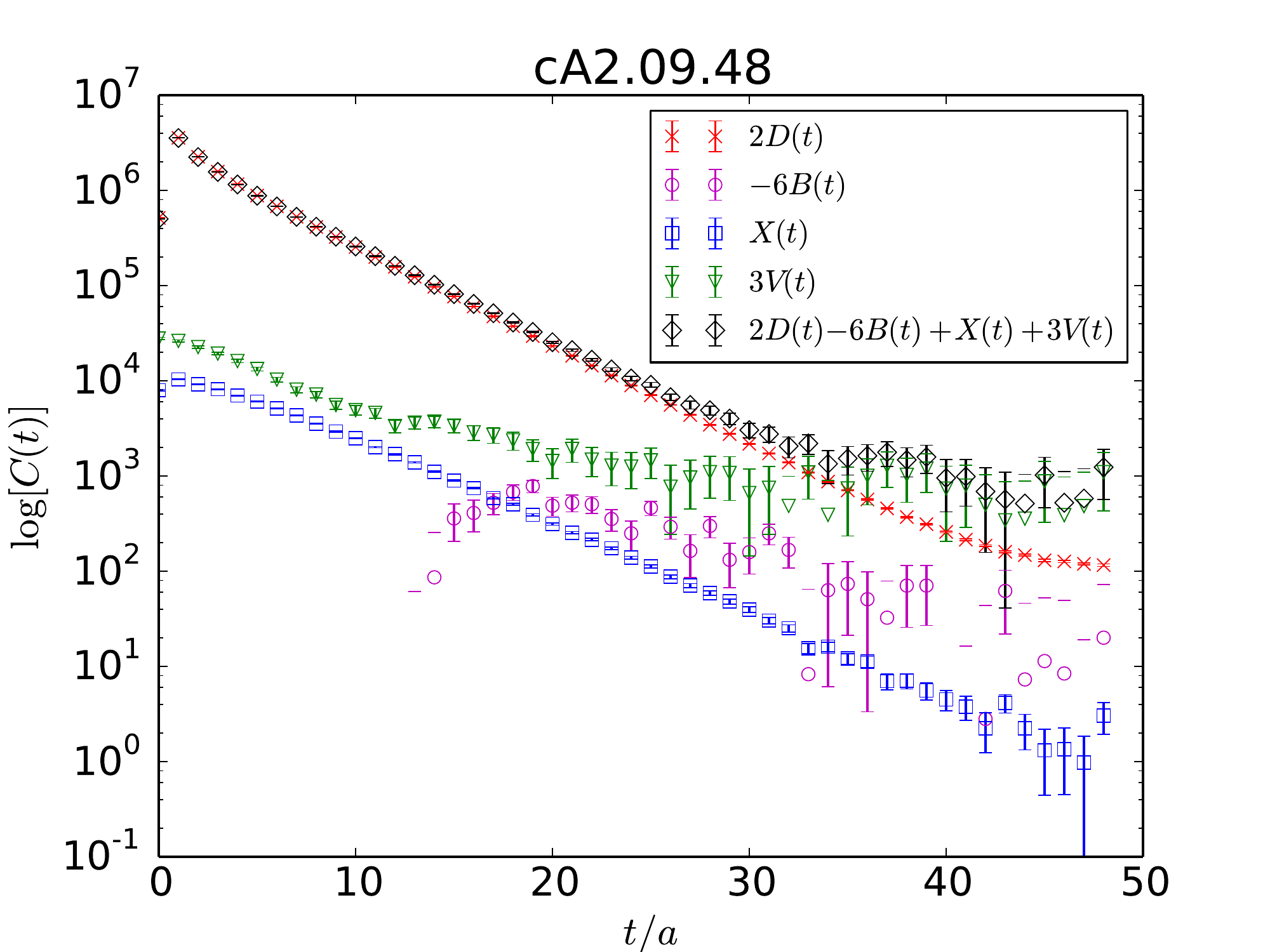}\includegraphics[width=0.5\linewidth]{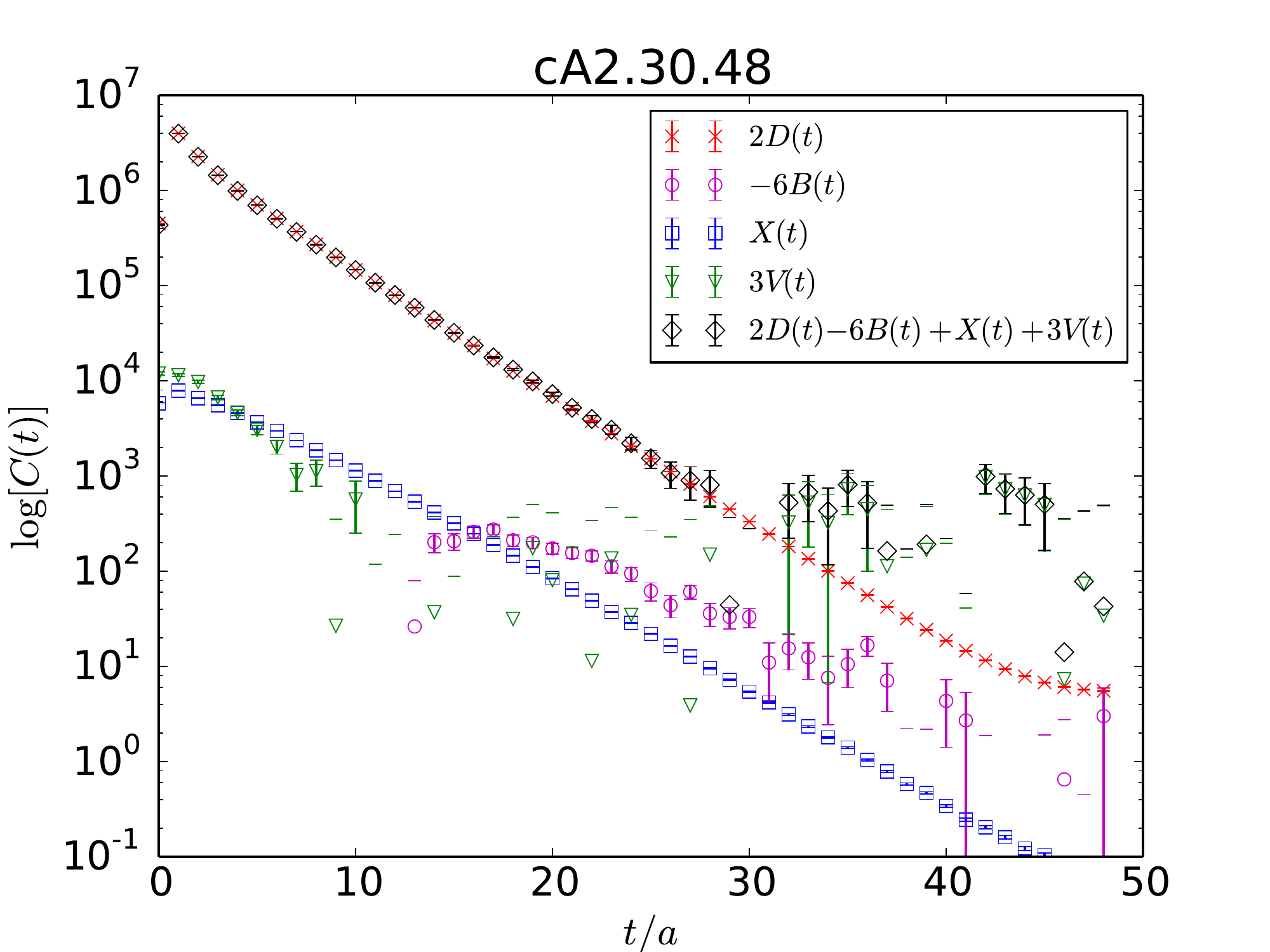} 
  \includegraphics[width=0.5\linewidth]{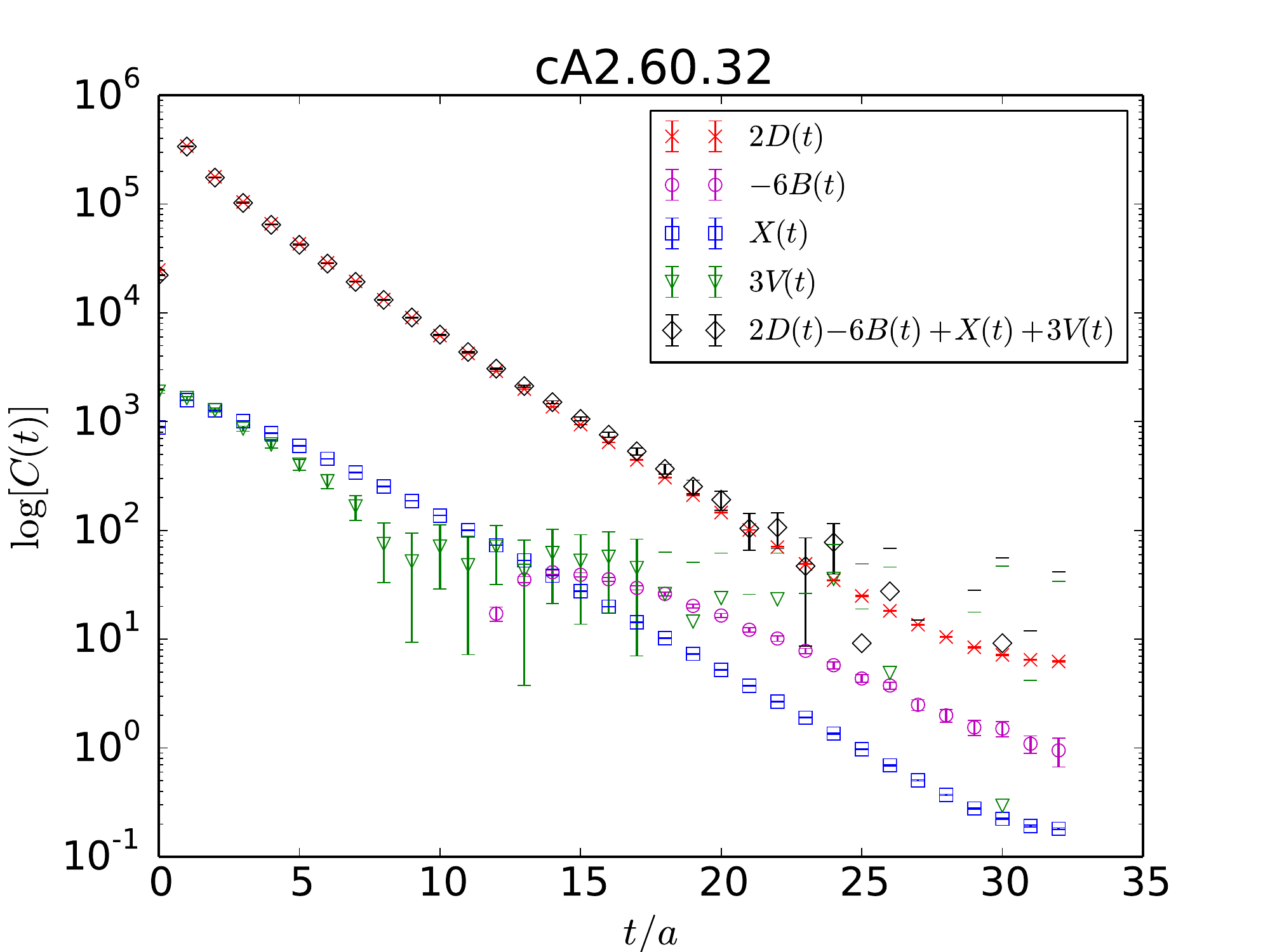} 
  \caption{Correlation functions of the operator
    $\mathcal{O}_{\pi\pi}^\mathrm{I=0}$ and the single diagrams
    $D,X,B,V$ for the three ensembles 
    listed in Table~\ref{tab:setup}.}
  \label{fig:corr}
\end{figure}

The energy of the isospin-0 $\pi\pi$ state can be computed from the
exponential decay in time of the correlation function
\begin{equation}
  \label{eq:Cpipi}
  C_{\pi\pi}(t) = \frac{1}{T}\sum_{t_{src}=0}^{T-1}\langle
  \mathcal{O}_{\pi\pi}^\mathrm{I=0} (t + t_{src})\  ( 
  \mathcal{O}_{\pi\pi}^\mathrm{I=0} )^\dagger (t_{src}) \rangle\,,
\end{equation}
where $T$ is the temporal lattice extend. The four diagrams
contributing to this correlation function, namely the direct connected
diagram $D(t)$, the cross diagram $X(t)$, the box diagram $B(t)$ and
the vacuum diagram $V(t)$, are depicted in
Fig.~\ref{fig:diagrams}\,(a)-(d). The correlation function can be 
expressed in terms of all relevant diagrams as 
\begin{equation}
 C_{\pi\pi}(t) = 2D(t) + X(t) - 6B(t) + 3V(t).
\end{equation}
$C_{\pi\pi}$ and the contributions from individual diagrams $D,X,B$
and $V$ are plotted in Fig.~\ref{fig:corr} for the three 
ensembles.

Even though we have full SU$(2)$ isospin symmetry in the valence
sector when using OS valence quarks as described above, we have to
consider effects of unitarity breaking. This may in particular happen
due to the vacuum diagram $V(t)$. There is no symmetry available
to prevent this diagram to couple for instance to intermediate states
of $n\geq 1$ unitary neutral pions (the neutral pion has vacuum
quantum numbers in maximally twisted mass lattice QCD), since parity
is not a good quantum number for our action. 
Since the neutral pion is the lightest meson in the spectrum
with Wilson twisted mass fermions at finite value of the lattice spacing,
the appearance of such states with $n=1$ (and maybe $n=2$) will dominate
the large Euclidean time behavior of the correlation function
$C_{\pi\pi}$ 
, if the overlap of the used interpolating operators
with these states is nonzero. In order to resolve this mixing, we build
a $2\times 2$ matrix of correlation functions 
\begin{equation}
  C_{ij}(t) = \frac{1}{T}\sum_{t_{src}=0}^{T-1} \langle
  \mathcal{O}_{i}(t + t_{src})
  \mathcal{O}_{j}^\dagger (t_{src}) \rangle
\end{equation}
with $i,j$ labeling the operator $\mathcal{O}_{\pi\pi}^\mathrm{I=0}$
and the unitary neutral pion operator 
\begin{equation}
  \pi^{0, uni}(t) = \sum_\mathbf{x} \frac{1}{\sqrt{2}}( \bar{u}\gamma_5
  u\ - \  \bar{d^\prime}\gamma_5 d^\prime)(\mathbf{x},t)\, , 
\end{equation}
where $u$ and $d^\prime$ are the (unitary) Wilson clover twisted mass
{\it up} and {\it down} quarks. We use $d^\prime$ to distinguish it
from OS {\it down} quark in Eq.~\ref{eq:piop}. The twisted
mass {\it up} quark coincides with the OS {\it up} quark with our
matching scheme of the OS to the unitary action.  

The diagrams contributing to the correlation function of the unitary
neutral pion operator are depicted in Fig.~\ref{fig:diagrams}\,(e) -
(f). The two operators couple solely via the vacuum diagram, see
Fig.~\ref{fig:diagrams}\,(g).   

The computation of the disconnected diagrams,
e.g. Fig.~\ref{fig:diagrams}\,(c), (d), (f), and (g), requires the quark
propagator  from a time slice $t$ to the same time for every $t$. This
has been a challenge in lattice QCD for decades.  By using the
stochastic LapH quark smearing
method~\cite{Peardon:2009gh,Morningstar:2011ka}, we have all-to-all
propagators available. The disconnected 
diagrams can be computed directly from those propagators. 

We can reduce lattice artefacts
in the vacuum diagrams following the ideas worked out in
Ref.~\cite{Ottnad:2015hva}. In the continuum limit the difference
between $u_+$ ($d_+$) quarks corresponding to $D_\ell^+$ and $u_-$
($d_-$) quarks corresponding to $D_\ell^-$
vanishes~\cite{Frezzotti:2004wz}. Therefore, we can write 
\begin{displaymath}
  \begin{split}
    \mathcal{O}(a) & =\ \langle \bar u_+ d_+(x)\ \bar d_+
    u_+(y) - \bar u_- 
     d_-(x)\ \bar d_-  u_-(y)\rangle \\
     & =\tr\{S^+(x,y)\ \ S^+(y,x)\ \} - \tr\{S^-(x,y)\ \ S^-(y,x)\ \} \\
     &= \tr\{S^+(x,y)\ \ S^+(y,x)\ \} - \tr\{(S^+(x,y)\ \ S^+(y,x))^\dagger\ \} \\
    & =\ 2 i \, \im\, \tr\{S^+(x,y)\ \ S^+(y,x)\ \}\,,\\
  \end{split}
\end{displaymath}
where $S^\pm \equiv (D_\ell^\pm)^{-1}$ are the OS quark propagators
and we have used $(S^+)^\dagger= \gamma_5 S^- \gamma_5$. This
shows that the imaginary part of the loop needed in the contraction of
the vacuum diagram $V$ is a pure lattice artefact and we will drop it
in the computation. The same argument holds for the 
vacuum diagrams shown in Fig.~\ref{fig:diagrams}\,(f) and (g). 

\subsection{Determination of the energies}

Due to the finite temporal extend $T$ of the lattice, the correlation
functions of multiparticle operators are polluted by so-called
thermal states~\cite{Feng:2009ij}. In the case of interest here,  
there is a constant contribution to $C_{\pi\pi}(t)$ of the form
\begin{displaymath}
  \propto\ |\langle \pi^{\pm,0}\,|\,  \mathcal{O}_{\pi\pi}^{I=0} \,
  |\,  \pi^{\pm,0} \rangle |^2 \cdot e^{- \mpios \,T}\,, 
\end{displaymath}
which vanishes in the infinite volume limit $T\to\infty$. However, at
finite $T$-values it will dominate the correlation function at large
Euclidean time.  To remove this artefact we define a shifted
correlation matrix 
\begin{equation}
  \label{eq:Ctilde}
  \tilde{C}(t) = C(t) - C(t+\delta t)\,.
\end{equation}
The new matrix $\tilde{C}$ is then free of any constant pollution 
from the thermal states. The value of $\delta t$ can be adjusted for
optimal results. We take $\delta t = 4$ in our analysis. Note that the
shifting procedure also subtracts any constants stemming from vacuum
expectation values.

The energy levels can then be obtained by solving the generalized
eigenvalue problem (GEVP)~\cite{Michael:1982gb}
 \begin{equation}
 \tilde{C}(t)\, v_n(t,t_0) = \lambda_n(t,t_0)\, \tilde{C}(t_0)\, v_n(t,t_0)\,.
 \end{equation}
The eigenvalues $\lambda_n(t,t_0)$ are expected to have the following
time dependence
\begin{equation}
\label{eq:sinhfit}
\lambda_n(t,t_0) = A_n \sinh \left[\left(\frac{T}{2} -  t -
  \frac{\delta t}{2}\right) E_n\right]\,. 
\end{equation}
The $\sinh$-like behavior comes from the shifting of the correlation
functions in Eq.~\ref{eq:Ctilde}. The energies $E_n$ are then obtained
by fitting the above functional form to the eigenvalues
$\lambda_n(t,t_0)$ in the range where the effective energy shows a
plateau. The value of $t_0$ should be chosen such that 
the correlation function at $t_0$ is dominated by the states we are
interested in. We tried various $t_0$ values 
in the range of 1 to 7 and found negligible differences in the
energies. In the following we set $t_0=1$.
With the two operators used here, we obtain two energy  
levels, one of which is far below the other one. The lower one agrees
with the unitary neutral pion mass, while the higher one is close to
$2 \mpios$. Hence, we identify the higher one to be the isospin-0 $\pi\pi$
state. In principle, multi neutral and charged unitary pion states
could also appear in the spectrum.
 To resolve these states, more operators need to
be included. We have tried so and merely found increased
statistical errors of the $I=0$ $\pi\pi$ state. Therefore, we cannot
finally exclude possible contaminations from such states.

\begin{figure}[t]
  \centering
 \includegraphics[width=0.5\linewidth]{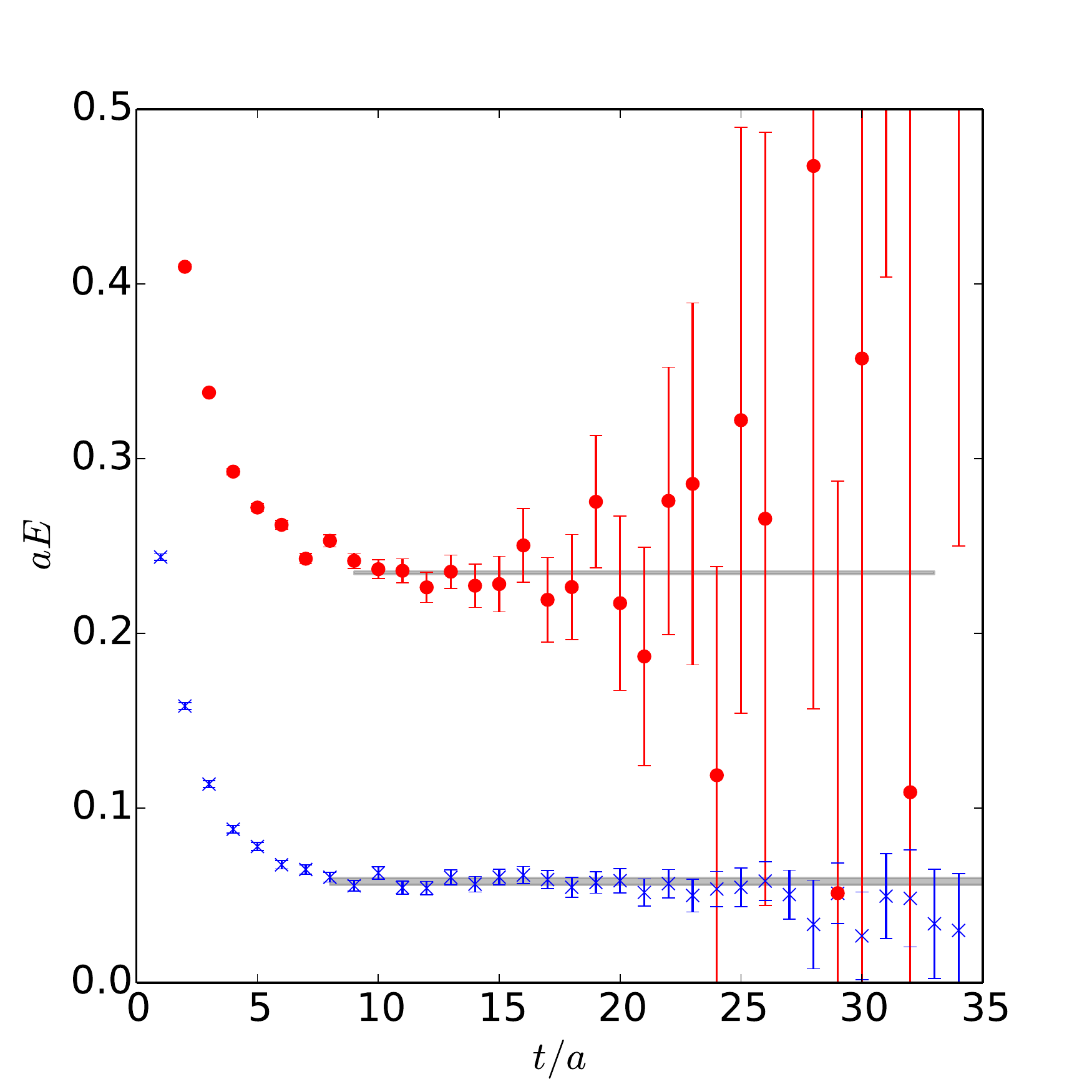}\includegraphics[width=0.5\linewidth]{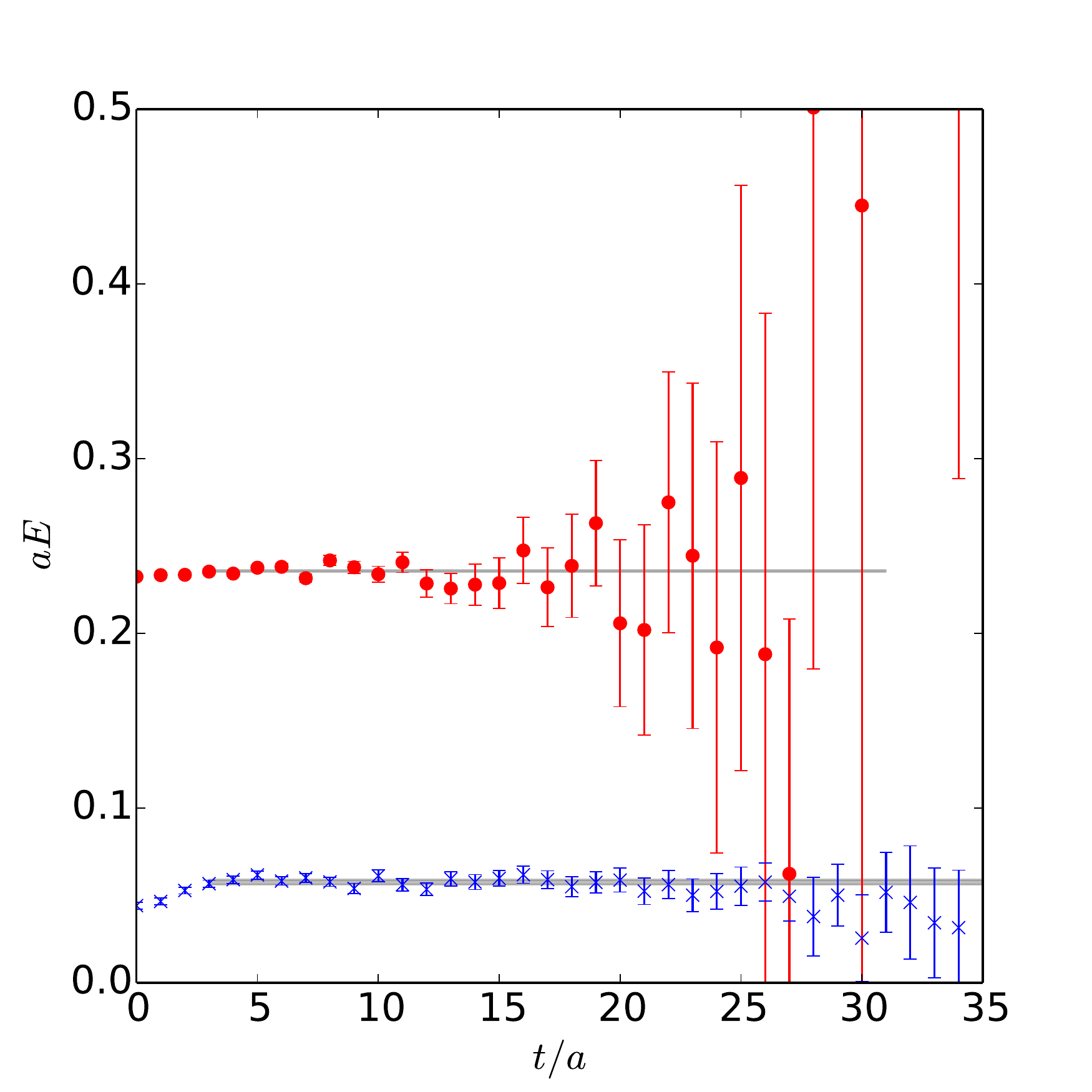}
  \caption{Effective energies for the ensemble cA2.09.48. The grey bars indicate the fitted values of
  the energies and the fit ranges. The left and right panels correspond to before and after removing the excited state contaminations, respectively. }
   \label{fig:eff_mass}
\end{figure}

As an example, the effective energies of the two eigenvalues for
ensemble cA2.09.48 are shown in Fig.~\ref{fig:eff_mass}\,(a). The
fitted energies and fit ranges are indicated by the grey bands in the
plot. 

To further improve our results we adopt a method to remove the excited
state contaminations ~\cite{Neff:2001zr}, which we have recently used
successfully to study $\eta$ and $\eta^\prime$
mesons~\cite{Jansen:2008wv,Michael:2013gka}. It is based on the 
assumption that vacuum diagrams are only sizeable for low lying
states, but negligible for higher excited states. Of course, the
validity of this assumption must be checked in the Monte-Carlo data.
In our case we know there is a very sizable disconnected contribution
to the unitary neutral pion, which represents a pure lattice
artefact~\cite{Dimopoulos:2009qv}. A similar contribution has not been
found to any other unitary correlation function. For the $\pi\pi$
correlation function there are indications that the disconnected
contribution is already small by itself~\cite{Fu:2013ffa}.

The connected contractions contributing to $\tilde{C}$ are computed
with sufficient precision, so we can reliably determine the
ground states in the connected correlators and subtract the
excited state contributions. We then build a correlation
matrix $\tilde{C}^\mathrm{sub}$ from the subtracted connected and the
original disconnected correlators. If disconnected contributions were
relevant only for the ground states, one should find -- after
diagonalizing  $\tilde{C}^\mathrm{sub}$ -- a plateau for both states
from small values of $t$ on. Note that with this procedure only the
small $t$ behavior of the correlation functions is altered.

To be more specific, the connected contributions to the correlation
function $C_{\pi\pi}(t)$ is given by
\begin{equation}
 C_{\pi\pi}^\mathrm{con}(t) = 2D(t) + X(t) - 6B(t)\,. 
\end{equation}
We fit the functional form Eq.~\ref{eq:sinhfit} to the shifted
correlator $\tilde{C}_{\pi\pi}^\mathrm{con}(t) =
C_{\pi\pi}^\mathrm{con}(t) -C_{\pi\pi}^\mathrm{con}(t-\delta t)$.
After obtaining the parameters $A$ and $E_{\pi\pi}^\mathrm{con}$ from
the fit, the connected correlator is reconstructed as 
\begin{equation}
  \tilde{C}_{\pi\pi}^\mathrm{con, sub}(t) = A\ \sinh
  \left[\left(\frac{T}{2} -  t - \frac{\delta t}{2}\right)
    E_{\pi\pi}^\mathrm{con}\right]\,, 
\end{equation}
in which the excited state contaminations are subtracted. We repeat the fit to
the data of $\tilde{C}_{\pi\pi}^\mathrm{con}(t)$ for many different
fit ranges. The expectation values of the fit parameters are computed
as the weighted median over these many fits~\cite{Helmes:2015gla}. By
doing this, the systematics arising from different fit ranges is
expected to be taken into account. The full correlator is then given by
$\tilde{C}_{\pi\pi}^\mathrm{sub}(t) = \tilde{C}_{\pi\pi}^\mathrm{con,
  sub}(t) + 3\tilde{V}(t)$, where $\tilde{V}(t)$ is the shifted vacuum
correlator $\tilde{V}(t) = V(t) - V(t+\delta t)$. The same procedure
is performed for the unitary $\pi^0$ correlation function.  

In Fig.~\ref{fig:eff_mass}\,(b), we present the effective energies of
the two eigenvalues of the subtracted correlator matrix
$\tilde{C}^\mathrm{sub}$ for the same ensemble as in
Fig.~\ref{fig:eff_mass}\,(a). Clearly a plateau appears at much earlier
$t$-value compared to Fig.~\ref{fig:eff_mass}\,(a), while the fitted
energies agree very well. Therefore, we use this procedure -- which
allows us to determine in particular the interacting energy $E_{\pi\pi}$
with much higher accuracy -- for the results presented here. 

\begin{figure}[t]
  \centering
    \includegraphics[width=0.5\linewidth]{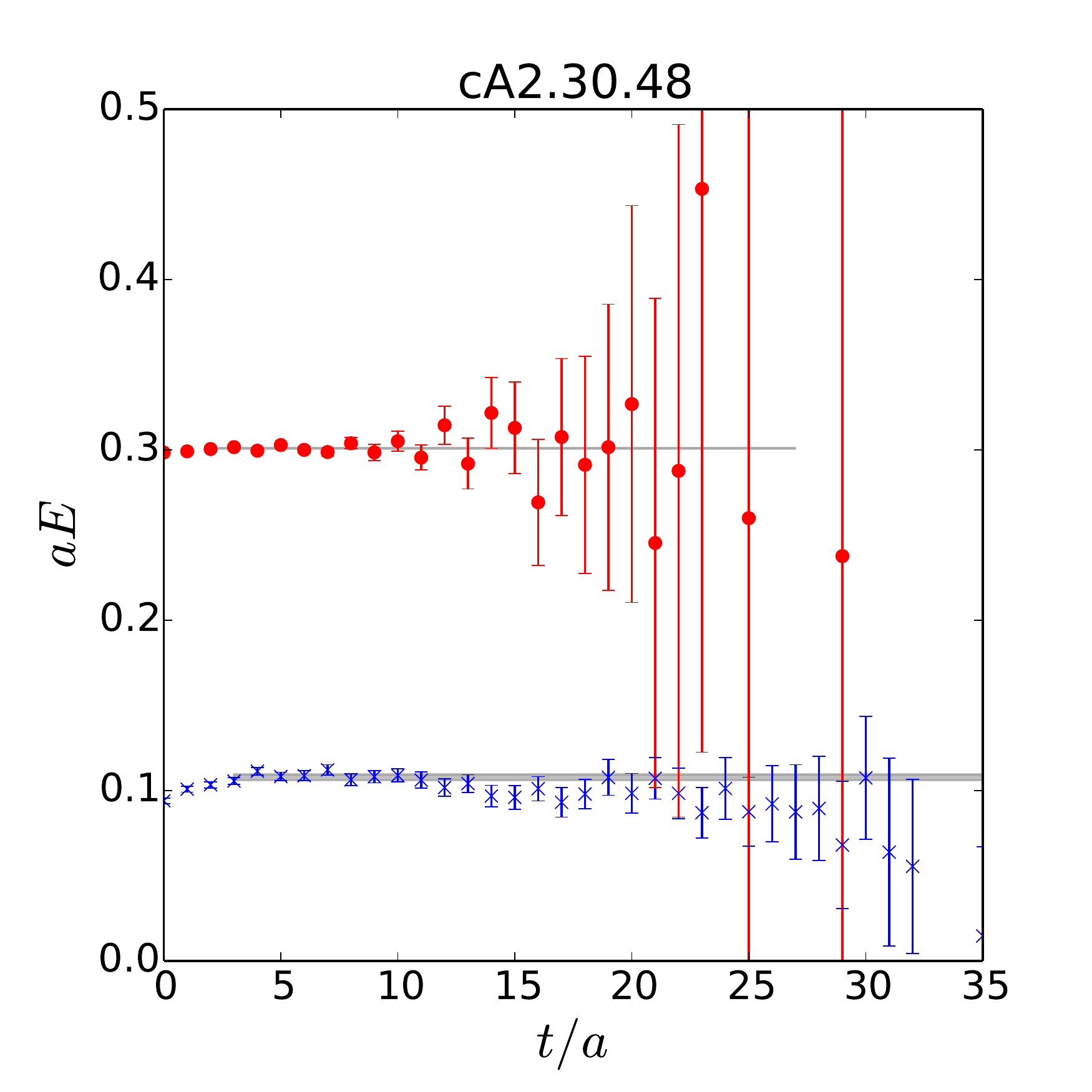}\includegraphics[width=0.5\linewidth]{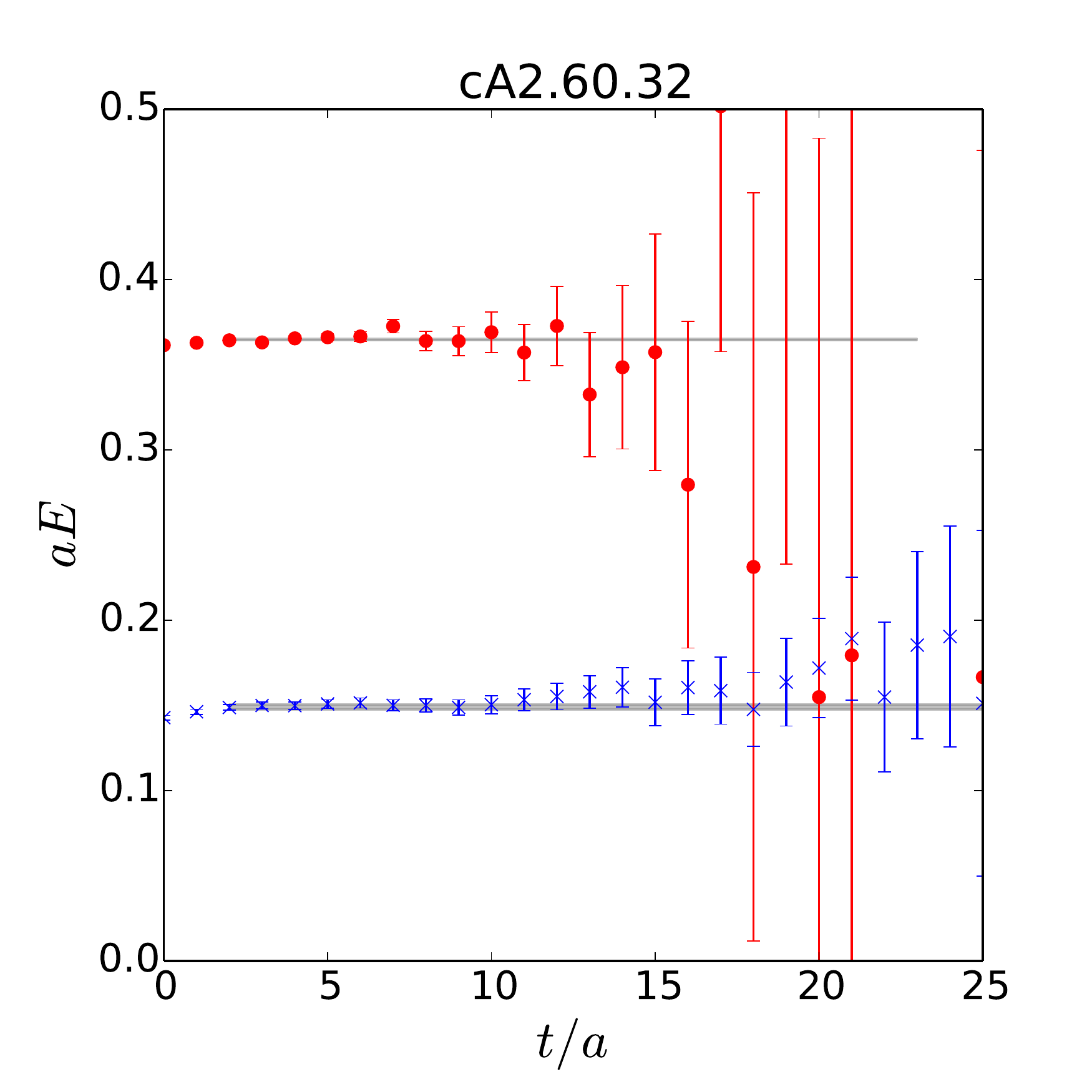} 
  \caption{Effective energies after removing the excited states
    contaminations for the ensembles $cA2.30.48$ and $cA2.60.32$. }
  \label{fig:eff_mass1}
\end{figure}

The effective energies after removing the exited states for the
ensembles  $cA2.30.48$ and $cA2.60.32$ are shown in
Fig.~\ref{fig:eff_mass1}. In Table~\ref{tab:EnergyResults}, we collect
the values of $E_{\pi\pi}$ and the unitary $\pi^0$ mass $\mpin$
obtained from the procedure described above. The unitary charged pion
mass $\mpi$ and the OS pion mass $\mpios$ are added for convenience.

In order to estimate possible artefacts from the mixing with the
unitary $\pi^0$, we also determined the energy $\hat{E}_{\pi\pi}$ by
fitting to only the single correlator
$\tilde{C}_{\pi\pi}^\mathrm{sub}(t)$, without including the operator
for the unitary neutral pion. The values are given in the last column of
Table~\ref{tab:EnergyResults}. One can see that the mean value of
$\hat{E}_{\pi\pi}$ is slightly lower than $E_{\pi\pi}$ for all three
ensembles, but they agree very well with each other
considering the statistical error. Keeping in mind that $\pi^0$ mixing
is purely a lattice artefact, the agreement between $E_{\pi\pi}$ and
$\hat{E}_{\pi\pi}$ indicates that we are not likely to suffer severe
lattice artefact here. This can be also seen in the small mass
splitting between the unitary charged and neutral pions.  

\begin{table}[t!]
  \centering
  \begin{tabular*}{\textwidth}{@{\extracolsep{\fill}}lccccc}
    \hline\hline
    Ensemble & $a \mpi$  & $a \mpin$  & $a \mpios$ & $a E_{\pi\pi}$ &$a \hat{E}_{\pi\pi}$ \\
    \hline
    $cA2.09.48$ &0.06212(6) &0.058(2) &0.11985(15) &0.2356(4) &0.2350(4)\\
    $cA2.30.48$ &0.11197(7) &0.108(2) &0.15214(11) &0.3010(3) &0.3009(3)  \\
    $cA2.60.32$ &0.15781(15) &0.149(2) &0.18844(24) &0.3647(5) &0.3645(5)\\
    \hline\hline
  \end{tabular*}
  \caption{Pion masses and the $\pi\pi$ interacting energies in lattice units for the
    three ensembles.}
  \label{tab:EnergyResults}
\end{table}


\section{Results}
\label{sec:results}
\subsection{Scattering length}
\label{subsec:ScatLen}

\begin{table}[t!]
 \centering
 \begin{tabular*}{\textwidth}{@{\extracolsep{\fill}}lccccc}
  \hline\hline
  Ensemble   & $(ak)^2$  & $ak \cot\delta(k)$ & $\frac{1}{2} a r_0 k^2$ & $\mpios a_0^\mathrm{I=0}$ &$\mpios/f_\pi^\mathrm{OS}$ \\
  \hline
    $cA2.09.48$  &-0.00049(4) &0.168(19) &0.0037(3)(2) &0.730(83)(1) &1.86(2)\\
  $cA2.30.48$    &-0.00050(4)  &0.167(19) &0.0042(3)(2) &0.94(11)(0) &2.21(1) \\
  $cA2.60.32$  &-0.00224(9) &0.074(7) &0.0224(9)(22) &-- &2.63(1) \\
  \hline\hline
 \end{tabular*}
 \caption{The values of squared scattering momentum $k^2$, $k
   \cot\delta(k)$, $\frac{1}{2}r_0 k^2$ (see appendix), $\mpios
   a_0^\mathrm{I=0}$ and $\mpios/f_\pi^\mathrm{OS}$ for the three considered
   ensembles. Values for $k^2$, $k\cot\delta$ and $r_0k^2$ are in
   lattice units. The first error is the statistical error, the second
   error comes from the uncertainty of $r_0$, see Appendix~\ref{app:EffRange}. } 
 \label{tab:results}
\end{table}

The scattering momentum $k^2$ is calculated from
Eq.~\ref{eq:ScatMomentum} with the energies $E_{\pi\pi}$ and the OS
pion masses listed in Table~\ref{tab:EnergyResults}. Then the
scattering length can be obtained from
Eq.~\ref{eq:ScatLen}. Considering the relatively strong interaction in
the isospin-0 $\pi\pi$ channel, one has to investigate the
contribution of $\mathcal{O}(k^2)$ and higher order terms in the
effective range expansion. Since we only have one energy level for
each pion mass, we are not able to determine the effective range $r_0$
with our lattice simulations. We rely on the $r_0$ values determined
from $\chi$PT~\cite{Gasser:1983yg}. See Appendix~\ref{app:EffRange}
for the details of the $r_0$ values used in our analysis.  

The values of $k^2$, $k\cot\delta(k)$ and $\frac{1}{2} r_0 k^2$ in
lattice units for all three ensembles are given in
Table~\ref{tab:results}. For the ensembles $cA2.09.48$ and
$cA2.30.48$ the scattering momentum is small due to the large
volume. Therefore, the contribution of $\frac{1}{2}r_0 k^2$ is
expected to be small. As visible from Table~\ref{tab:results}, the 
value of $\frac{1}{2} r_0 k^2$ is indeed less than $3\%$ of $k\cot
\delta(k)$ for these two ensembles.  We compute the scattering length
from Eq.~\ref{eq:ScatLen} by ignoring the $\mathcal{O}(k^4)$ term,
which is well justified for the ensembles $cA2.09.48$ and
$cA2.30.48$. The values of $\mpios a_0^\mathrm{I=0}$ for these two
ensembles are also given in Table~\ref{tab:results}. The first error
is the statistical error and the second error comes from the
uncertainty of the effective range $r_0$.  

As for the ensemble cA2.60.32, the value of $\frac{1}{2} r_0 k^2$ is
rather large compared to $k\cot \delta(k)$. This indicates that the
effective range expansion up to $\mathcal{O}(k^2)$ might not be a good
approximation and the $\mathcal{O}(k^4)$ term might not be
negligible. Since the contribution of $\mathcal{O}(k^4)$ is unclear,
we refrain from giving the scattering length for this ensemble. There
are two possible reasons for the invalidity of the effective range
expansion. First, due to the relatively small volume of the ensemble
cA2.60.32, the value of $k^2$ is not small enough to make the
expansion converge rapidly. Second, at the pion mass around 400~MeV,
which is the OS pion mass of the ensemble cA2.60.32, 
there might be virtual or bound state poles appearing in the isospin-0
$\pi\pi$ scattering
amplitude~\cite{Albaladejo:2012te,Pelaez:2010fj,Hanhart:2008mx,hanhart:priv,Bernard:2010fp,Doring:2011vk}.  
The appearance of such states will give a very large scattering length
-- positively (negatively) large if it was a virtual (bound) state. Hence,
the leading order in the effective range expansion,
i.e. $\frac{1}{a_0}$, becomes very small compared to the higher
orders. In this case the NLO term $\frac{1}{2}r_0
k^2$ can contribute significantly compared to the LO term
even when $k^2$ is small. Assuming that the contribution of
$\mathcal{O}(k^4)$ term is not bigger than the $\mathcal{O}(k^2)$
term, we can qualitatively estimate the scattering length for this
ensemble to be a very large positive number, which features a virtual
state. However, we do not exclude the possibility of a bound state
because we do not include single meson operators when we compute the
matrix of correlation functions. Including these operators might  
change the resulting spectrum and thus the scattering length.

\subsection{Chiral extrapolation}

\begin{figure}[t]
  \includegraphics[width=0.6\linewidth]{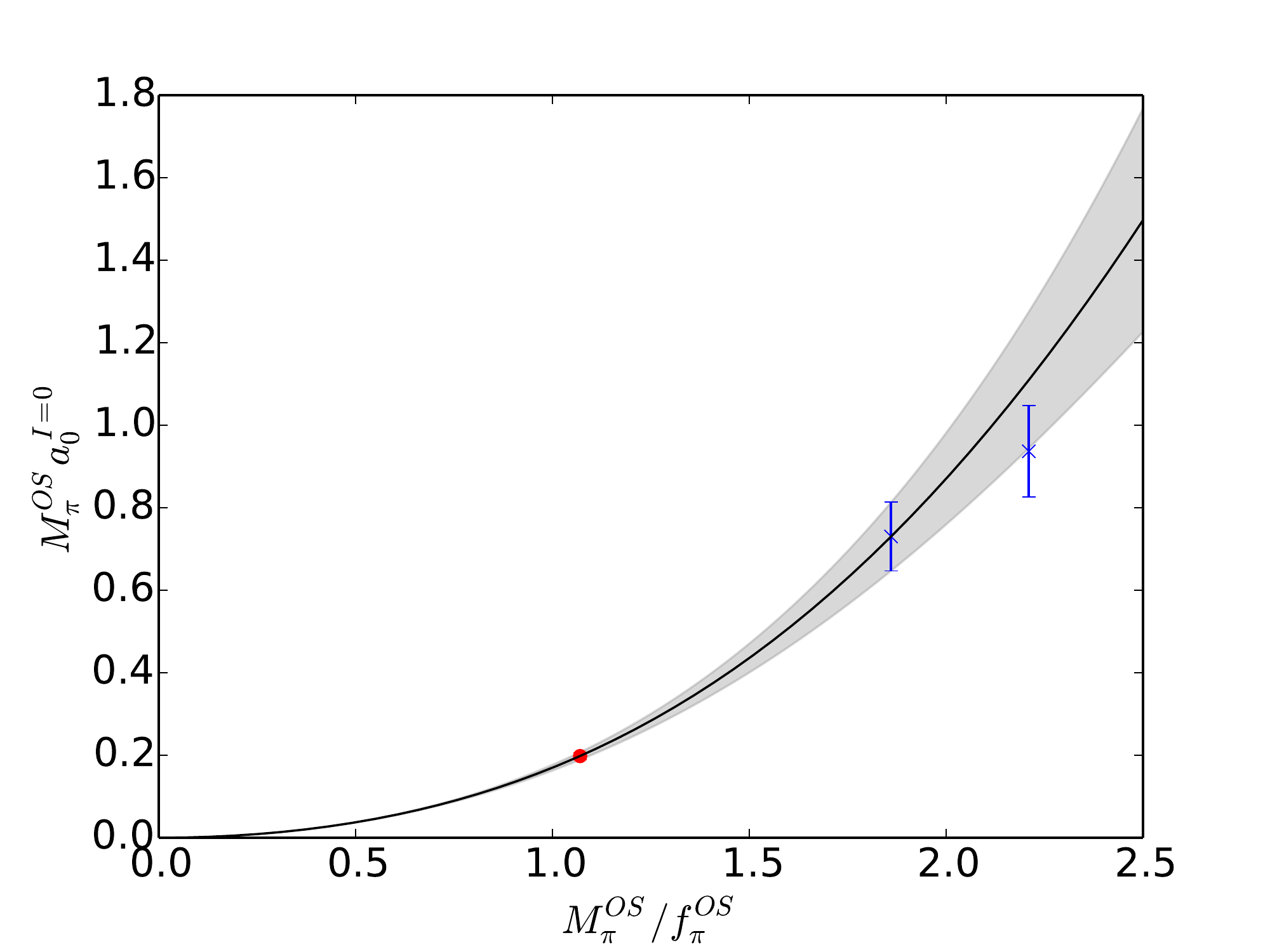} 
  \caption{Chiral extrapolation using only the data point with lower pion mass.
  The grey band represents the uncertainty. The red point indicates the extrapolated
  value at physical pion mass. }
  \label{fig:Chiral_fit}
\end{figure}

In order to obtain the scattering length at the physical pion mass, one
needs to perform a chiral extrapolation. $\pi\pi$ scattering has been
studied extensively in $\chi$PT in the
literature~\cite{Weinberg:1966kf, Gasser:1983yg, Bijnens:1997vq,
  Bijnens:1995yn, Colangelo:2001df}. Since we only have
two data points, we fit the NLO $\chi$PT formula, which contains one free
parameter, to our data. When expressed in terms of $M_\pi$ and 
$f_\pi$ computed from lattice simulations, the formula
reads~\cite{Fu:2013ffa} 
\begin{equation}
  \label{eq:chipt}
  M_\pi a_0^\mathrm{I=0}\ = \ \frac{7 M_\pi^2}{16 \pi f_\pi^2} \left[ 1 - \frac{
      M_\pi^2}{16\pi^2
      f_\pi^2}\left(9\ln\frac{M_\pi^2}{f_\pi^2}-5-\ell^\mathrm{I=0}_{\pi\pi}\right)\right]\,,
\end{equation}
where $\ell^\mathrm{I=0}_{\pi\pi}$ is a combination of the low energy
coefficients $\bar{l}_i$'s :
\begin{equation}
  \ell^\mathrm{I=0}_{\pi\pi} = \frac{40}{21} \bar{l}_1 + \frac{80}{21}
  \bar{l}_2 - \frac{5}{7} \bar{l}_3 + 4 \bar{l}_4 + 9 \ln
  \frac{M_\pi^2}{f_{\pi, phy}^2} \,. 
\end{equation}
In this expression, the renormalization scale is set to be the
physical pion decay constant $f_{\pi, phy}$. Note that we work in the
normalization with $f_{\pi, phy}=130.4\ \mathrm{MeV}$. 
 By using the formula in Eq.~\ref{eq:chipt}, we have ignored the effects of unitarity breaking. 
In principle, we should use mixed action $\chi$PT to perform the chiral extrapolation. The $\chi$PT
for the mixed action with twisted mass sea quarks and OS valence quarks is presented in Ref.~\cite{WalkerLoud:2009nf}.
However, using the two data points at one value of lattice spacing, we are not able to implement the mixed action $\chi$PT formula. With this caveat in mind, we proceed with our analysis.

The OS pion decay constant $f_{\pi}^\mathrm{OS}$ has not been determined by ETMC yet.
We compute $f_{\pi}^\mathrm{OS}$ for the three ensembles used in this
paper. The values of $\mpios/f_{\pi}^\mathrm{OS}$  are collected in the last
column of Table~\ref{tab:results}. The details of their computation are
presented in Appendix~\ref{app:OSfpi}. We recall that the OS pion mass
values are larger than their unitary counterpart, see
Table~\ref{tab:EnergyResults}, such that our lowest OS pion mass value
is at around 250~MeV. 

The method we are applying here is valid only in the elastic
region. Therefore, the pion mass values must be small enough to be
below threshold where the $\sigma$ meson becomes stable. This
threshold is not known exactly, but results obtained with the 1-loop
inverse amplitude method~\cite{Hanhart:2008mx} (see
also Refs.~\cite{Pelaez:2010fj,Hanhart:2014ssa, Albaladejo:2012te}) 
suggest that $M_\pi < 400\ \mathrm{MeV}$ should be safe~\cite{hanhart:priv}.
Our two data points are obtained at pion mass around 250~MeV and 320~MeV 
respectively, both are below this threshold. 
Furthermore, the pion mass value should also be small enough to make the 
chiral expansion valid. To be safe, we perform the chiral extrapolation 
using only the data point with the lower pion mass (~250~MeV). The results 
of this extrapolation are given in Table~\ref{Table:fitresults} as
fit-1 and illustrated in Fig.~\ref{fig:Chiral_fit}. The results of the  
fit with the two data points, which are given
in Table~\ref{Table:fitresults} as fit-2, agree with fit-1 within errors.
We take the difference as an estimate of
the systematics arising from chiral extrapolation. This leads to our 
final result for the scattering length:
\begin{equation}
M_\pi a^\mathrm{I=0}_0 = 0.198(9)_\mathrm{stat}(6)_\mathrm{sys}\,.
\end{equation}
We remark here that the extrapolation is strongly constrained since
$\mpios a_0^{I=0}$ must vanish in the chiral limit. This explains the
small statistical uncertainty on the value extrapolated to the
physical point.
\begin{table}
  \begin{tabular*}{\linewidth}{@{\extracolsep{\fill}}lrr}
    \hline\hline
     &fit-1   & fit-2  \\
         \hline\hline
    $M_\pi a^\mathrm{I=0}_0$(Phy.)   & 0.198(9) &0.192(5) \\
    $\ell_{\pi\pi}^\mathrm{I=0}$     & 30(8)  & 24(4)  \\
    $\chi^2/\mathrm{dof}$            & --    & 0.75/1\\
    \hline\hline
  \end{tabular*}
   \caption{Results of the NLO chiral fit. fit-1 includes only one data
     point, namely cA2.09.48, while fit-2 includes both, cA2.09.48 and
     cA2.30.48.}
  \label{Table:fitresults}
\end{table}

We compare our result in Table~\ref{tab:CompMpia0} to other results
available in the literature. Our result for $M_\pi a_0^{I=0}$ is
lower, but within errors still compatible with most recent
experimental, lattice and Roy equations results. Due to
our comparably low value for $M_\pi a_0^{I=0}$, the value for
$\ell_{\pi\pi}^{I=0}$ is also relatively low. This is a direct
consequence of the NLO $\chi$PT description we are using.

\begin{table}[t!]
  \centering
  \begin{tabular*}{\linewidth}{@{\extracolsep{\fill}}lrr}
    \hline\hline
    & $M_\pi a_0^\mathrm{I=0}$ & $\ell_{\pi\pi}^\mathrm{I=0}$ \\
    \hline\hline
    This work & $0.198(9)(6)$ & $30(8)(6)$ \\
    Fu~\cite{Fu:2013ffa} & $0.214(4)(7)$ & $43.2(3.5)(5.6)$ \\

    \hline
    Weinberg~\cite{Weinberg:1966kf} & $0.1595(5)$ & $-$ \\
    CGL~\cite{Colangelo:2001df} & $0.220(5)$ & $48.5(4.3)$\\

    \hline
    NA48/2~\cite{Batley:2010zza} & $0.220(3)(2)$ & $$ \\
    E865~\cite{Pislak:2003sv} & $0.216(13)(2)$ & $45.0(11.2)(3.5)$ \\
    \hline\hline
  \end{tabular*}
  \caption{Comparison of results available in the literature for $M_\pi
    a_0^\mathrm{I=0}$ and $\ell_{\pi\pi}^\mathrm{I=0}$.} 
  \label{tab:CompMpia0}
\end{table}


\section{Discussion and Summary}

In this paper, the isospin-0 $\pi\pi$ scattering is studied with
L\"uscher's finite volume formalism in twisted mass lattice  
QCD. We use a mixed action approach with the OS action in the valence
sector to circumvent the complications arising from isospin
symmetry breaking in the twisted mass quark action. 
The stochastic LapH quark smearing method is used to compute all-to-all quark
propagators, which are required to compute the quark disconnected
diagrams contributing  to the isospin-0 $\pi\pi$ correlation
function. The lowest energy level in the rest frame is extracted for
three $N_f =2$ ensembles with three different pion mass values. The
scattering length is computed with L\"uscher's formula for the two
ensembles with the lowest pion mass values. For the third ensemble
with the largest pion mass value the scattering length cannot be
determined reliably. In the computation of the scattering length we
include the $\mathcal{O}(k^2)$ term in the effective range expansion
using values for the effective range, which we compute using $\chi$PT.
The chiral extrapolation of $M_\pi a^{I=0}_0$ is
performed using NLO $\chi$PT. Extrapolated to the physical value of
$M_\pi/f_\pi$, our result is $M_\pi a^{I=0}_0 = 0.198(9)(6)$, which is
compatible within errors with the newer experimental and theoretical
determinations available in the literature. 

Since we work at a single lattice spacing value only, we cannot
estimate lattice artefacts in our result. In particular, we cannot
exclude that our result is affected by residual systematic
uncertainties stemming from unitarity breaking, which will vanish in
the continuum limit. Moreover, the control over higher order
contributions from $\chi$PT is rather limited. We cannot exclude
that such contributions are sizable. 

For these reasons a future study should include several lattice
spacing values and ideally ensembles at the physical point. In order
to avoid isospin breaking and unitarity breaking effects, we will
repeat this computation with an action without isospin breaking.

\begin{acknowledgments}

We thank the
members of ETMC for the most enjoyable collaboration. The computer
time for this project was made available to us by the John von
Neumann-Institute for Computing (NIC) on the Jureca and Juqueen
systems in J{\"u}lich. 
We thank A.~Rusetsky and Zhi-Hui Guo for very useful discussions and R.~Brice\~no for
valuable comments. We are grateful to Ulf-G. Mei\ss ner for carefully reading this manuscript and helpful comments.
This project was funded by the DFG as a project in
the Sino-German CRC110. S.~B. has received funding from the Horizon 2020 research and innovation 
program of the European Commission under the Marie Sklodowska-Curie
programme GrantNo. 642069. This work was granted access to the HPC resources IDRIS under the
allocation 52271 made by GENCI.
The open source software
packages tmLQCD~\cite{Jansen:2009xp}, Lemon~\cite{Deuzeman:2011wz}, DD$\alpha$AMG~\cite{Alexandrou:2016izb}
and
R~\cite{R:2005} have been used.
\end{acknowledgments}

\appendix

%
\section{Effective range from $\chi$PT} 
\label{app:EffRange}
In order to investigate the contribution of the $\mathcal{O}(k^2)$ term in the effective range expansion, we need to know the value of the effective range $r_0$. As explained in Section~\ref{subsec:ScatLen}, we estimate $r_0$ from $\chi$PT. 

In Ref.~\cite{Gasser:1983yg}, the chiral expansion of the threshold parameter $b_0^0$ to NLO is given as
\begin{equation}
b_0^0 = \frac{1}{2 \pi f_{\pi}^2} \left\{ 1+\frac{M_\pi^2}{f_\pi^2}\left[ - \frac{13}{12\pi^2} \ln \frac{M_\pi^2}{\mu^2}  + 32 l_1^r + 24 l_2^r + 4 l_4^r - \frac{13}{96 \pi^2} \right] \right\},
\end{equation}
where $\mu$ is the renormalization scale, $l_1^r$, $l_2^r$ and $l_4^r$ are the renormalized, quark mass independent couplings. In this expression, we have replaced the low energy parameters $M$ and $F$ with their (lattice) physical values $M_\pi$ and $f_\pi$ using the NLO chiral expansions of $M_\pi^2$ and $f_{\pi}$, which are given in the same reference. Please also note that our convention of $f_{\pi}$ ($\sim$130~MeV) is different from the $F_{\pi}$ ($\sim$92.4~MeV) used in Ref.~\cite{Gasser:1983yg}. 

The effective range $r_0$ is related to $b_0^0$ as $r_0 = - 2 b_0^0 M_\pi$. In order to avoid the uncertainty in lattice scale setting, we write $r_0$ in lattice units as a function of the dimensionless parameters $aM_\pi$ and $M_\pi^2/f_\pi^2$ : 
\begin{displaymath}
  \begin{split}
\frac{r_0}{a} & =  - \frac{2 b_0^0 M_\pi^2}{a M_\pi} \\ 
                    & =   - \frac{1}{aM_\pi} \frac{M_\pi^2}{\pi f_\pi^2}\left\{ 1+\frac{M_\pi^2}{f_\pi^2}\left[ - \frac{13}{12\pi^2} \ln \frac{M_\pi^2}{f_\pi^2}  + 32 l_1^r + 24 l_2^r + 4 l_4^r - \frac{13}{96 \pi^2} \right] \right\}. \\
  \end{split}
\end{displaymath}
Here the renormalization scale $\mu$ is set to be the physical pion decay constant $f_\pi^\mathrm{phy}$. To write the formula as a function of $M_\pi/f_\pi$, we have replaced $f_\pi^\mathrm{phy}$ with $f_\pi$. The corrections due to this replacement appear at next-to-next-to-leading order. 

We take the values of the scale independent couplings $\bar{l}_1$, $\bar{l}_2$ and $\bar{l}_4$ determined in Ref.~\cite{Bijnens:2014lea}: 
\begin{equation}
\label{eq:lbars}
\bar{l}_1 = -0.4 \pm 0.6, \quad \bar{l}_2 = 4.3 \pm 0.1,  \quad \bar{l}_4 = 4.3 \pm 0.3.
\end{equation}
From the relations between $l_i^r$ and $\bar{l}_i$ 
\begin{equation}
l_i^r = \frac{ \gamma_i}{32\pi^2} \left( \bar{l}_i + \ln \frac{M_\pi^2}{\mu^2} \right)
\end{equation}
with $\gamma_1 = 1/3$, $\gamma_2=2/3$ and $\gamma_4 = 2$, we calculate the values of $l_i^r$ at $\mu = f_\pi^\mathrm{phy}$:
\begin{equation}
\label{eq:lir}
l_1^r = -0.0003(6) ,\quad  l_2^r=0.0094(2) ,\quad l_4^r=0.0281(19) .
\end{equation}

The effective range is calculated with the $l_i^r$'s in Eq.~\ref{eq:lir} and the values of $a\mpios$ and $M_\pi^\mathrm{OS}/f_\pi^\mathrm{OS}$ in Table~\ref{tab:EnergyResults} and Table~\ref{tab:results}. The results of $r_0/a$ for the three ensembles are presented in Table~\ref{tab:r0}. The errors are estimated from the errors of $l_i^r$'s and the statistical uncertainties of $a\mpios$ and $M_\pi^\mathrm{OS}/f_\pi^\mathrm{OS}$. 

\begin{table}[t!]
 \centering
 \begin{tabular*}{\textwidth}{@{\extracolsep{\fill}}lcccc}
  \hline\hline
  Ensemble & cA2.09.48  & cA2.30.48  & cA2.60.32  \\
  \hline
  $r_0/a$ &-14.9(0.8) &-17(1)  &-20(2) \\
   \hline\hline
 \end{tabular*}
 \caption{The effective range in lattice unit.}
 \label{tab:r0}
 \end{table}

\section{Determination of the OS $f_\pi$ values}
\label{app:OSfpi}
The chiral extrapolation of the $I=0$ scattering length is most
conveniently performed in $M_\pi/f_\pi$. For this reason we need to
compute also the OS pion decay constant. It is given by the following
relation~\cite{Constantinou:2010gr}
\begin{equation}
  f_\pi^\mathrm{OS} = Z_A \frac{\langle 0| A_0|\pi\rangle}{\mpios}\,,
\end{equation}
with the (OS valence quark) axial vector component $A_0=\bar u \gamma_5\gamma_0 d$. 
The corresponding renormalization constant $Z_A$ has been determined
in Ref.~\cite{Alexandrou:2015sea} for the action and $\beta$-value
used here. It reads 
\[
Z_A\ =\ 0.7910(6)\,.
\]
The matrix element $\langle 0| A_0|\pi\rangle$ can be determined from
suitable correlation functions. We used the operator
\[
\mathcal{O}_A\ =\ \sum_{\mathbf{x}}
\bar u\gamma_5\gamma_0 d(\mathbf{x},t)
\]
together with $\pi^+$ from Eq.~\ref{eq:piop} to build a $2\times2$
correlation matrix. For the matrix element we need local operators,
hence, we cannot use sLapH. Instead, we performed dedicated inversions
with local operators and the one-end-trick~\cite{Boucaud:2008xu}.
Since the off-diagonal correlators have a $\sinh$-like behaviour, we
perform a constrained fit to this correlator matrix to determine the
ground state energy and the corresponding matrix element at large
Euclidean times.

\bibliographystyle{h-physrev5}
\bibliography{bibliography}

\newpage 
\begin{appendix}
\end{appendix}

\end{document}